\documentclass[12pt]{amsart}
\usepackage{amsmath,amssymb,amsthm,verbatim,url,color,geometry, pdflscape, bigstrut}
\usepackage{rotating}

\newenvironment{changemargin}[2]{
\begin{list}{}{\setlength{\topsep}{0pt}\setlength{\leftmargin}{0pt}
\setlength{\rightmargin}{0pt}\setlength{\listparindent}{\parindent}\setlength{\itemindent}{\parindent}
\setlength{\parsep}{0pt plus 1pt}\addtolength{\leftmargin}{#1}\addtolength{\rightmargin}{#2}}\item }{\end{list}
}

\long\def\forget#1\forgotten{}
\newcommand{\nc}{\newcommand}
\nc{\transpose}[1]{#1^\mathrm{t}}
\nc{\V}{{\{0,1\}^*}}
\nc{\Impl}{\Rightarrow}
\nc{\rev}{\mathrm{r}}
\newtheorem{thm}{Theorem}[section]\nc{\bthm}{\begin{thm}} \nc{\ethm}{\end{thm}}
\newtheorem{prop}[thm]{Proposition}\nc{\bprp}{\begin{prop}} \nc{\eprp}{\end{prop}}
\newtheorem{fact}[thm]{Fact}\nc{\bfct}{\begin{fact}} \nc{\efct}{\end{fact}}
\newtheorem{prob}[thm]{Problem}\nc{\bprb}{\begin{prob}} \nc{\eprb}{\end{prob}}
\newtheorem{lem}[thm]{Lemma}\nc{\blem}{\begin{lem}} \nc{\elem}{\end{lem}}
\newtheorem{claim}[thm]{Claim}\nc{\bclm}{\begin{claim}} \nc{\eclm}{\end{claim}}
\newtheorem{cor}[thm]{Corollary}\nc{\bcor}{\begin{cor}} \nc{\ecor}{\end{cor}}
\newtheorem{conj}[thm]{Conjecture}\nc{\bcnj}{\begin{conj}} \nc{\ecnj}{\end{conj}}
\theoremstyle{definition}\newtheorem{defn}[thm]{Definition}\nc{\bdfn}{\begin{defn}} \nc{\edfn}{\end{defn}}
\theoremstyle{remark}
\newtheorem{rem}[thm]{Remark}\nc{\brem}{\begin{rem}}\nc{\erem}{\end{rem}}
\newtheorem{cnv}[thm]{Convention}\nc{\bcnv}{\begin{cnv}} \nc{\ecnv}{\end{cnv}}
\newtheorem{exam}[thm]{Example}\nc{\bexm}{\begin{exam}}\nc{\eexm}{\end{exam}}
\newtheorem{alg}[thm]{Algorithm}\nc{\balg}{\begin{alg}}\nc{\ealg}{\end{alg}}
\nc{\bpf}{\begin{proof}}\nc{\epf}{\end{proof}}\nc{\be}{\begin{enumerate}}\nc{\ee}{\end{enumerate}}
\nc{\bi}{\begin{itemize}}\nc{\itm}{\item}\nc{\ei}{\end{itemize}}
\nc{\Cayley}{\op{Cayley}}
\nc{\x}{\times}
\nc{\inv}{^{-1}}
\nc{\cT}{\mathcal{T}}
\nc{\cK}{\mathcal{K}}
\nc{\cD}{\mathcal{D}}
\nc{\tr}{\op{tr}}
\nc{\sm}{\setminus}
\nc{\sub}{\subseteq}
\nc{\set}[2]{\left\{#1\,:\,#2\right\}}
\nc{\ceq}{\stackrel{\cdot}{=}}
\nc{\my}[1]{\textcolor{red}{\sf #1}}
\nc{\mx}[1]{\begin{pmatrix}#1\end{pmatrix}}
\nc{\op}[1]{\operatorname{#1}}
\nc{\SL}[1]{\SLtwo({#1})}
\nc{\bbF}{\mathbb{F}}
\nc{\SLtwo}{\op{SL}_2}
\nc{\SLq}{$\mathrm{{SL}}_2(\mathbb{F}_q)$}
\nc{\SLx}{\mathrm{{SL}}_2(\mathbb{F}_q[x])}
\nc{\TZ}{Tillich--Z\'{e}mor}
\nc{\GL}{\mathrm{GL}}
\nc{\bM}{\begin{smallmatrix}}
\nc{\eM}{\end{smallmatrix}}
\nc{\ed}{\end{document}}

\title[$\SLtwo$ homomorphic hash]{$\SLtwo$ 
homomorphic hash functions: Worst case to average case reduction and 
short collision search}
\author{Ciaran Mullan}
\email{}
\address{Technische Universit\"at Darmstadt, Fachbereich Informatik, Kryptographie und Computeralgebra,
Hochschulstra\ss{}e 10, 64289 Darmstadt, Germany}

\author{Boaz Tsaban}
\address{Department of Mathematics, Bar Ilan University, 5290002 Ramat Gan, Israel
and
Faculty of Mathematics and Computer Science, Weizmann Institute of Science, 7610001 Rhovot, Israel}
\email{tsaban@math.biu.ac.il}
\urladdr{http://www.cs.biu.ac.il/~tsaban}

\keywords{$\SLtwo$ hash, homomorphic hash function, Cayley hash function, Tillich--Z\'emor hash, expander graphs}

\subjclass[2010]{
94A60,          
20G40.          
}

\begin{document}

\forget
We study homomorphic hash functions into SL(2,q), the 2x2 matrices with determinant 1 over the
field with $q$ elements. 
Modulo a well supported number theoretic hypothesis, which holds in particular for concrete
homomorphisms proposed thus far, we provide a worst case to average case reduction for these hash functions:
upto a logarithmic factor, a random homomorphism is as secure as _any_ concrete homomorphism.
For a family of homomorphisms containing several concrete proposals in the literature,
we prove that collisions of length O(log(q)) can be found in running time O(sqrt(q)).
For general homomorphisms we offer an algorithm that, heuristically and according to experiments,
in running time O(sqrt(q)) finds collisions of length O(log(q)) for q even, and length O(log^2(q)/loglog(q))$ for arbitrary q.
While exponetial time, our algorithms are faster in practice than all earlier generic algorithms,
and produce much shorter collisions.
\forgotten

\begin{abstract}
We study homomorphic hash functions into $\SL{q}$, the $2\x 2$ matrices with determinant $1$ over the
field with $q$ elements. 
Modulo a well supported number theoretic hypothesis, which holds in particular for concrete
homomorphisms proposed thus far, we provide a worst case to average case reduction for these hash functions:
upto a logarithmic factor, a random homomorphism is as secure as \emph{any} concrete homomorphism.
For a family of homomorphisms containing several concrete proposals in the literature,
we prove that collisions of length $O(\log{q})$ can be found in running time $O(\sqrt{q})$.
For general homomorphisms we offer an algorithm that, heuristically and according to experiments,
in running time $O(\sqrt{q})$ finds collisions of length $O(\log q)$ for $q$ even, and length $O({\log}^2{q}/\log{\log{q}})$ for arbitrary $q$.
While exponetial time, our algorithms are faster in practice than all earlier generic algorithms,
and produce much shorter collisions.
\end{abstract}

\maketitle

\section{Introduction}

Let $\V$ be the monoid of all finite bitstrings with string concatenation as monoid multiplication and the empty string as identity
element.
Let $\SL{q}$ be the group of $2\x 2$ matrices of determinant $1$ with entries in the finite field $\mathbb{F}_{q}$ with $q=p^n$ elements.
Over 20 years ago, Z\'emor~\cite{Zemor91} proposed a general hash function construction employing homomorphisms $h\colon \V\to\SL{q}$, that is, functions $h$ with the property that $h(uv)=h(u)h(v)$ for all $u,v\in \V$.
For a pair of elements $A=(A_0,A_1)$ of $\SL{q}$, denote by $h_A$ the unique homomorphism such that $h_A(0)=A_0$ and $h_A(1)=A_1$. A bitstrings $b_1\ldots b_m\in \V$ is hashed to the matrix
$$h_A(b_1\ldots b_m) = h_A(b_1)\cdots h_A(b_m) = A_{b_1}\cdots A_{b_m}\in \SL{q}.$$
Variations of Z\'emor's original scheme were proposed in, e.g., \cite{TZ2, CLG}. 
We refer to the survey of Petit and Quisquater~\cite{PQRubik} for an introduction to this family 
of hash functions and its features. 

At present, feasible cryptanalyses on this construction apply only for very special instances of $A$ and $q$. 
A very efficient cryptanalysis for the case where $q$ is a power of $2$ and $A$ is a specific, 
natural pair of matrices
was recently provided by Grassl et al.~\cite{GIMS11};
see the survey~\cite{PQRubik} and the paper~\cite{BSV}
for a discsussion of the known cryptanalytic results and their limitations.

We study Z\'emor's construction in its full generality. 
Based on a well supported conjecture concerning expander graphs, 
in Section~\ref{sec:random} we prove that $\SL{q}$ homomorphic hash functions 
based on a random homomorphism is as secure as any concrete homomorphism, 
upto a logarithmic factor in collision length.
Such \emph{worst case to average case reductions}, 
also called \emph{random self-reducibility}, are very desirable in cryptographic 
primitives, see, e.g., Ajtai's seminal paper \cite{Ajtai} and the numerous works 
that cite it. This puts Z\'emor's construction at the frontiers of provably 
secure hash functions and motivates a further study of this approach.

The running time of all algorithms studied in this paper is measured by the number of multiplications of elements of $\SL{q}$.
In Section~\ref{sec:rational} we provide an algorithm producing, modulo the same well-known conjecture, collisions of length $O(\log{q})$ in time $O({\sqrt{q}})$, for arbitrary $q$ and a class of
homomorphisms including those in \cite{TZ1} (end of $\S{6}$, $i=2$), \cite{Zemor91}, \cite{TZ2}, and \cite{AbdKim98}.
In Section~\ref{sec:generic}, for random $(A_0,A_1)$ and arbitrary $q$ we provide a collision search algorithm,
and show, heuristically, that it finds collisions of length $O({\log}^2{q}/\log{\log{q}})$ in running time $O(\sqrt{q})$.
In Section \ref{sec:even} we show that, for messages of all practical sizes, our algorithm is faster and produces much shorter collisions than the
best known subexponential time algorithm due to Faug\`{e}re et al.~\cite{Faugere}.
Moreover, it is shown that the heuristic methods of Petit~\cite{PetitTowards} and Faug\`{e}re et al.\ can be used, for $q$ a power of $2$, to reduce an \emph{arbitrary} pair of generators $(A_0,A_1)$ into a form in which
our algorithm of Section \ref{sec:rational} applies. Consequently, we obtain collisions of linear length for arbitrary
homomorphisms into $\SL{2^n}$.

The theory employed in  Sections \ref{sec:random} and \ref{sec:rational} may be used to obtain, in a rigorous manner,
estimations for the first phase of an earlier algorithm of Petit et al.~\cite{PQTZ09}. We survey this algorithm in Appendix \ref{apx:PQTZ}. For an optimal choice of parameters we estimate its performance, which turns out to be not as good as our new algorithms. Furthermore, our algorithms are conceptually simpler: unlike
Petit et al.\ we do not appeal to discrete logarithm solving or use of the LLL algorithm.
We remark that Petit et al.'s algorithm produces bitstrings hashing to the identity matrix, of length linear in $p$. While the same can be done with our first algorithm of Section 3, apparently this cannot be achieved with our second, more general algorithm of Section 4.

Finally, in Appendix \ref{apx:nopali} we prove that palindromic collisions, as exploited by Grassl et al.~\cite{GIMS11} in their efficient attack for $q$ even, do not exist for arbitrary $q$, based on the same natural generating sets.

We mention, here only, that the memory required by our algorithm can be made negligible, using distinguished points as
in \cite[$\S{6}$]{PQTZ09}.
All of our estimations are supported by extensive computer experiments. When we are interested in estimating
the involved constants, we use $\lg$, the logarithm in base $2$, instead of $\log$. The operator $|~|$ means: absolute
value when applied to a real number, cardinality when applied to a set, determinant when applied to a matrix,
and bitlength when applied to a bitstring.

\vspace{2mm} 

Hash functions typically fall into one of two categories:
a mathematical design with transparent security but slow performance,
or an ad hoc design and fast, but obscure security.
The study of Z\'{e}mor-like hash functions is worthy of investigation as 
it may lead to the design of a fast hash function whose security is based 
on a natural mathematical problem. Moreover, properties built into these 
constructions, including bit-level hashing (as opposed to fixed size blocks) 
and homomorphism and parallelism properties, may find use in applications.
Our results, including the cryptanalytic ones, suggest that random instances
of the studied hash family may meet the mentioned goals.
We hope that our new mathematical treatment and simpler collision-finding 
algorithms will encourage further research in this field.

\section{Worst case to average case reduction}\label{sec:random}

In earlier papers on $\SL{q}$ hash functions (see \cite{PQRubik} and references therein),
much effort has been put on selecting the pair $(A_0,A_1)=(h(0),h(1))$ carefully.
One motivation was to have the hash function efficiently implementable. Another was
to have it more ``secure'': that small differences in the hashed messages are detectable,
and that the hash function is ``mixing''.
Here, we show that hashing with a random homomorphism---that is, with a pair of random elements $(A_0,A_1)$---is
not less secure than hashing with any prescribed, carefully chosen homomorphism. The price may be at most a logarithmic
factor in the collision length.

In this paper, by \emph{graph} we always mean a directed one.
Let $G$ be a group.
For a generating subset $S$ of $G$, the \emph{Cayley graph}
of $(G,S)$ is the graph $\Gamma$ with $G$ as set of vertices, and an edge from $g$ to $ga$ for each $g\in G,a\in S$.
This is a regular graph of degree $|S|$.
A regular graph $\Gamma=(V,E)$ is an \emph{$\epsilon$-expander} if, for each set of vertices $U\sub V$ with $|U|\le |V|/2$,
the set $N(U)$---of neighbors of elements of $U$---satisfies $|N(U)\sm U|\ge \epsilon |U|$.
(Necessarily, $\epsilon\le 1$ in this case.)
Surveys on expander graphs are available in \cite{HLW06, Go08, Lub12}.

For a $d$-regular graph $\Gamma$ with adjacency matrix $A$, let
$$\lambda(\Gamma) = \max\set{|\lambda|}{\lambda\mbox{ is an eigenvalue of }A,\ |\lambda|\neq d}.$$
Throughout this section, $|G|$ should be thought of as tending to infinity, whereas $|S|$ (and thus $d$) and $\epsilon$ should be considered constant.
We will use the following known facts.\footnote{The references given are to the surveys, where the primary references can be found.}
In Item (2) of the following theorem, the vector ${\hat A}^m\mathbf{p}$ 
describes the distribution on $V$ corresponding to choosing a vertex according to the distribution $\mathbf{p}$, and then performing $m$ steps of
random walk on the graph, where in each step one moves to a uniformly chosen neighbor of the present vertex. (As
there are loops on the vertices, one may remain at the same vertex after the step.)

\bthm\label{thm:expgeneral}
Let $\Gamma=(V,E)$ be a finite $d$-regular graph.
\be
\itm If $\Gamma$ has loops on each vertex and $\Gamma$ is an $\epsilon$-expander,
then $d-\lambda(\Gamma)\ge \epsilon^2/(4+2\epsilon^2)$ \cite[Theorem E.7]{Go08}.
\itm Let $\alpha=\lambda(\Gamma)/d$ and $\hat A=\frac{1}{d}A$.
Let $\mathbf{u}$ be the uniform distribution on $V$,
and let $\mathbf{p}$ be an arbitrary distribution on $V$.
Then, for each event $B$:
$$\left|\Pr_{{\hat A}^m\mathbf{p}}[B]-\Pr_{\mathbf{u}}[B]\right|\le
\frac{1}{2}\|{\hat A}^m\mathbf{p}-\mathbf{u}\|_1\le
\frac{1}{2}\sqrt{|V|}\cdot \alpha^m$$
for all $m$ \cite[Theorem 3.2]{HLW06}.
\ee
\ethm

Let $G$ be a finite group, and let $g=(g_0,\dots,g_{k-1})$ be a $k$-tuple of generators of $G$.
The homomorphic hash function $h_g\colon \{0,\dots,k-1\}^*\to G$ is defined by
$$h_g(b_1 b_2\ldots b_m) := g_{b_1}g_{b_2}\cdots g_{b_m}\in G$$
for all $b_1 b_2\ldots b_m\in\{0,\dots,k-1\}^*$.
For a set $S\subseteq G$, define $S^{\pm 1}:=S\cup S^{-1}$, 
where $S^{-1}:=\set{s^{-1}}{s\in S}$.

The first item of the following proposition was pointed out to us by E. Breuillard.\footnote{We state and prove this observation
in a slightly more general setting than the one provided by Breuillard, but the argument is identical to Breuillard's.}

\bprp\label{prp:exp}
Let $G$ be a finite group, and let $S=\{g_0,\dots,g_{k-1}\}$ be generators of $G$ such that the Cayley graph of
$(G,S^{\pm 1})$
is an $\epsilon$-expander.
Then:
\be
\itm The Cayley graph of $(G,S)$ is an $\epsilon/(k+1)$-expander.
\itm Let $m=(c(k+1)^3/\epsilon^2)\log|G|$, $c>5/2$.
Let $\mathbf{u}$ be the uniform distribution on $G$.
Let $e$ be the neutral element of $G$, and set $g=(g_0,\dots,g_{k-1},e)$.
If $v\in\{0,\dots,k\}^m$ is chosen uniformly at random, then for each event $B$:
$$\left|\Pr[h_g(v)\in B]-\Pr_{\mathbf{u}}[B]\right|\le
\frac{1}{2}\|h_g(v)-\mathbf{u}\|_1 <
\frac{1}{2|G|^{c/5-1/2}}.$$
\ee
\eprp
\bpf
(1) Let $\delta=\epsilon/(k+1)$.
Assume that there is $U\sub G$ such that $|U|\le |G|/2$ and $|US\sm U|<\delta |U|$.
Fix $s\in S$. In particular, $|Us\sm U|<\delta |U|$, and thus
$$|Us\inv\cap U|=|U\cap Us\inv|=|(Us\cap U)s\inv|=|Us\cap U|\ge (1-\delta) |U|.$$
Thus, $|Us\inv\sm U|<\delta|U|$, and therefore
$$\epsilon|U|\le |US^{\pm1}\sm U|\le |US\inv\sm U|+|US\sm U|<k\delta|U|+\delta|U|=(k+1)\delta|U|=\epsilon|U|;$$
a contradiction.

(2) Let $\delta = \epsilon/(k+1)$.
By (1), the Cayley graph of $(G,S)$ is a $\delta$-expander.
The Cayley graph of $(G,S\cup\{e\})$, where $e$ is the neutral element of $G$,
is the Cayley graph of $(G,S)$, with a loop added at each vertex.
As $N(U)\sm U$ does not change when adding loops,
the Cayley graph of $(G,S\cup\{e\})$ is a $\delta$-expander, too.

As the Cayley graph $\Gamma$ of $(G,S\cup\{e\})$ has loops on all vertices,
Theorem \ref{thm:expgeneral} applies. As $\delta\le1/2$,
$$k+1-\lambda(\Gamma)\ge \frac{\delta^2}{4+2\delta^2}
> \frac{\delta^2}{5}.$$
Thus,
$$\frac{\lambda(\Gamma)}{k+1}<1-\frac{\delta^2}{5d}=1-\frac{\epsilon^2}{5d^3}.$$
Let $v=b_1 b_2\ldots b_m\in\{0,\ldots,k\}^m$
be chosen uniformly at random. Then
$$h_g(v)=g_{b_1}\cdots g_{b_m}$$
is the endpoint of a uniform random walk of length $m$ in the
Cayley graph $\Gamma$ of $(G,S\cup\{e\})$, starting at $e$.
By Theorem \ref{thm:expgeneral}, for $\alpha = 1-\epsilon^2/5(k+1)^3$:
$$\left|\Pr[h_g(v)\in B]-\Pr_{\mathbf{u}}[B]\right|\le
\frac{1}{2}\|h_g(v)-\mathbf{u}\|_1 <
\frac{1}{2}\sqrt{|G|}\cdot \alpha^m.$$
Let $m=c/5\cdot\log_{1/\alpha}|G|$. Then
$$\sqrt{|G|}\cdot\alpha^m=\sqrt{|G|}\cdot\alpha^{c/5\cdot \log_{1/\alpha}|G|}=
\sqrt{|G|}(|G|^{\log_{1/\alpha}\alpha})^{c/5} =\sqrt{|G|}\cdot|G|^{-{c/5}}=
1/|G|^{c/5-1/2}.$$
As
$$\log\alpha=\log(1-\epsilon^2/5(k+1)^3)<-\epsilon^2/5(k+1)^3,$$
we have that
$$\log_{1/\alpha}|G|=\frac{\log|G|}{\log\frac1\alpha}=
\frac{\log|G|}{-\log\alpha}<
\frac{\log|G|}{\epsilon^2/5(k+1)^3}=
\frac{5(k+1)^3}{\epsilon^2}\log|G|,
$$
and $m$ is as required.
\epf

As its proof indicates, the following theorem can be generalized to arbitrary, not necessarily equal, numbers
of given generators and random elements. We state it, though, in the form needed here.

\bthm\label{thm:exp}
Let $G$ be a finite group, and let $g=(g_0,g_1)$ be a pair of generators of $G$ such that the Cayley graph of
$(G,\{g_0^{\pm 1},g_1^{\pm 1}\})$ is an $\epsilon$-expander.
Assume that if $r=(r_0,r_1)\in G^2$ is chosen uniformly at random,
one can find in time $O(t)$, with non-negligible probability, collisions of length $O(l)$ in $h_r$.
Then one can find with the same probability and time $O(t)$,
collisions in the original hash function $h_g$, of length $O(l\log |G|/\epsilon^2)$.
\ethm
\bpf
Let $m=(c\cdot3^3/\epsilon^2)\log|G|$, with $c$ large enough (say, $10$).
Let $g=(g_0,g_1,e)$.
Take uniformly random, independent $v_0,v_1\in\{0,1,2\}^m$.
By Proposition \ref{prp:exp}, $r_0:=h_g(v_0)$ and $r_1:=h_g(v_1)$ are
statistically indistinguishable from independent, uniformly random elements of $G$.
A collision
$$h_{r}(b_1b_2\cdots b_{l_1})=h_{r}(c_1c_2\cdots c_{l_2})$$
of length $l:=\max\{l_1,l_2\}$ yields the collision
$$h_g(v_{b_1}v_{b_2}\cdots v_{b_{l_1}})=h_g(v_{c_1}v_{c_2}\cdots v_{c_{l_2}})$$
of length $O(ml)=O(l \log|G|/\epsilon^2)$.
As $e$ is the neutral element of $G$, this is also a (typically, shorter) collision of length $O(ml)$ in the original
generators $g_0$ and $g_1$.
\epf

Let $\epsilon>0$. Let $\mathbb{P}$ be a family of prime powers. For each $q\in\mathbb{P}$,
assume that $A_0^{(q)},A_1^{(q)}\in\SL{q}$ are generators such that the Cayley graph of
$(\SL{q},\{A_0^{(q)},A_1^{(q)}\}^{\pm 1})$ is an $\epsilon$-expander. Then, by Theorem \ref{thm:exp},
the associated hash functions $h_{A^{(q)}}$ are not more secure than random hash functions $h\colon\{0,1\}^*\to\SL{q}$.
In other words, the hash functions $h_R$ with $R=(R_0,R_1)\in\SL{q}^2$ a uniformly random pair of matrices are the strongest in terms of collision resistance.

This observation is applicable in our setting for two reasons. The first is that, in all concrete proposals
made thus far (e.g., \cite{TZ1,TZ2,Zemor91}) the corresponding Cayley graph was proved to be an expander.
The second, more general, is the following well known and well supported conjecture (cf.\ Conjecture 2.9 in \cite{Lub12}).

\bcnj[Lubotzky]\label{cnj:expander}
There is a constant $\epsilon>0$ such that, for all prime powers $q$, and all generators $A_0,A_1$ of $\SL{q}$, the Cayley graph of $(\SL{q},\{A_0^{\pm 1},A_1^{\pm 1}\})$ is an $\epsilon$-expander.
\ecnj

In the case where the generators $A_0,A_1$ are chosen at random and $q$ is prime, this conjecture was proved to hold
for randomly chosen matrices, with probability going to $1$ as $q$ increases, by Bourgain and Gamburd \cite{BG08}.
Breuillard, Green, Guralnick and Tao \cite{BGGT} have recently extended this result
to $q$ an arbitrary prime power.
From another direction, Breuillard and Gamburd \cite{BG10} proved that there is a set of primes $q$,
of density $1$ in the primes, for which the conjecture holds regardless of the choice of generators.

\section{Collisions of linear length}\label{sec:rational}

The following theorem provides an algorithm for finding collisions of length $O(\log q)$ in
time $O(\sqrt{q})$, for a special class of generators. This class includes a substantial portion of the
concrete pairs of generators proposed in the literature, including the ones in
\cite{TZ1} (end of \S{6}, $i=2$), \cite{Zemor91}, \cite{TZ2}, and \cite{AbdKim98}.
According to Lubotzky's above-mentioned Conjecture \ref{cnj:expander} and the discussion
following it, $\epsilon$ may be viewed as a constant in the following theorem.

\bthm\label{thm:lincol}
Let $A=(A_0,A_1)$ be a pair of generators of $\SL{q}$ such that $|A_0-A_1|=0$.
If the Cayley graph of $(\SL{q},\{A_0^{\pm 1},A_1^{\pm 1}\})$ is an $\epsilon$-expander, then
a collision on $h_A$ of length $O(\log q/\epsilon^2)$ can be found in time $O(\sqrt{q})$.
\ethm

The remainder of this section details the proof of Theorem \ref{thm:lincol}.
Let
$$\cT:=\set{\mx{\alpha & \beta\\ 0 & \alpha\inv}}{0\neq \alpha\in\bbF_q,\ \beta\in\bbF_q}$$
be the subgroup of $\SL{q}$ consisting of all upper triangular matrices.

\blem\label{lem:rational}
For generators $A_0,A_1$ of $\SL{q}$, the following conditions are equivalent:
\be
\itm $|A_0-A_1|=0$.
\itm There exists $P\in\SL{q}$ and $\xi_0,\xi_1\in\bbF_q$ such that
$$P\inv A_i P=\mx{\xi_i & -1\\ 1 & 0}$$
for $i=0,1$.
\ee
\elem
\bpf
$(1)\Impl (2)$: Let $v$ be a nontrivial vector with
$A_0v-A_1v=(A_0-A_1)v=\vec{0}$.
Let
$$u:=A_0v=A_1v.$$
Assume that $u=\alpha v$ for some $\alpha\in\bbF_q$.
Let $P\in\SL{q}$ be a matrix whose first column is $v$. Then
$$P\inv A_i P=\mx{\alpha & *\\ 0 & *}\in\cT$$
for $i=0,1$, and thus $A_0,A_1$ do not generate $\SL{q}$; a contradiction.

Thus, $u$ is linearly independent of $v$.
Let $Q$ be the matrix whose columns are $(-u,v)$ and let $P = |Q|^{-1}Q$.
Then
$$P\inv A_i P=\mx{* & -1\\ * & 0},$$
and having determinant $1$, we arrive at (2).

$(2)\Impl (1)$:
$$|A_0-A_1|=|P\inv (A_0- A_1) P|=|P\inv A_0P-P\inv A_1 P|=
\left|\mx{\xi_0-\xi_1 & 0\\ 0 & 0}\right|=0.\qedhere$$
\epf

By Lemma \ref{lem:rational}, we may assume that
$$A_i =\mx{\xi_i & -1\\ 1 & 0}$$
for $i=0,1$.

\bdfn
For a bitstring $v=b_1b_2\ldots b_m\in V$, we define $v^\rev:=b_m\ldots b_2b_1$ as the reversal bitstring.
\edfn

\blem\label{lem:hashvr}
Let $A=(A_0,A_1)$ be a pair of elements of $\SL{q}$ with
$$A_i=\mx{\xi_i & -1\\1 & 0}$$
for $i=0,1$.
For a bitstring $v$ let
$$\mx{\alpha & \beta \\ \gamma & \delta }:=h_A(v).$$
Then
$$h_A(v^\rev)=\mx{\alpha & -\gamma\\ -\beta & \delta}.$$
\elem
\bpf
By induction on $|v|$.
If $|v|=1$ then $h_A(v)$ is $A_0$ or $A_1$, both of the desired form.
Assume the result holds for $v$. Then for each $i\in\{0,1\}$, we have by the induction hypothesis that
\begin{eqnarray*}
h_A(vi) & = & \mx{  \alpha & \beta \\ \gamma & \delta }\mx{\xi_i & -1\\1 & 0} =
\mx{\alpha\xi_i+\beta & -\alpha\\ \gamma\xi_i+\delta & -\gamma},\\
h_A(iv^\rev ) & = & \mx{\xi_i & -1\\1 & 0}\mx{  \alpha & -\gamma \\     -\beta & \delta } =
\mx{\alpha\xi_i+\beta & -\gamma\xi_i-\delta\\ \alpha  & -\gamma}.
\end{eqnarray*}
Thus, $h_A((vi)^\rev )=h_A(iv^\rev )$ has the desired form.
\epf

Let
$$\cK:=\set{\mx{1 & \beta\\ 0 & 1}}{\beta\in\bbF_q}.$$
$\cK$ is a subgroup of $\cT$.
Since $\cK$ is abelian, hashing into $\cK$ with two noncommuting bitstrings $u,v$ (i.e., such that $uv\neq vu$)
yields the collision
$$h_A(uv)=h_A(u)h_A(v)=h_A(v)h_A(u)=h_A(vu).$$


\bprp\label{prp:revcol}
Let $A=(A_0,A_1)$ be a pair of elements of $\SL{q}$, with
$$A_i=\mx{\xi_i & -1\\1 & 0}$$
for $i=0,1$.
Let $b_1b_2\ldots b_m$ be a bitstring such that $h_A(b_1b_2\ldots b_m)\in\cT$.
Then for all $i\in\{0,1\}$
$$h_A(ib_m\ldots b_2)\in\cT,$$
and
$$h_A(b_1b_2\ldots b_m ib_m\ldots b_2),h_A(ib_m\ldots b_2b_1b_2\ldots b_m)\in\cK.$$
\eprp
\bpf
Let $v=b_1b_2\ldots b_m$, and assume that
$$h_A(v)=\mx{\alpha & \beta\\ 0 & \alpha\inv}.$$
By Lemma \ref{lem:hashvr},
\begin{eqnarray*}
h_A(ib_m\ldots b_2) & = & h_A(i)h_A(b_m\ldots b_2)h_A(b_1)h_A(b_1)\inv\\
                    & = & h_A(i)h_A(b_m\ldots b_2b_1)h_A(b_1)\inv\\
                    & = & \mx{\xi_i & -1\\1 & 0}\mx{\alpha & 0\\ -\beta & \alpha\inv}\mx{\xi_{b_1} & -1\\1 & 0}\inv\\
                    & = & \mx{* & -\alpha\inv\\ \alpha & 0}\mx{0 & 1\\-1 & \xi_{b_1}}\\
                    & = & \mx{\alpha\inv & *\\ 0 & \alpha}\in\cT.
\end{eqnarray*}
Moreover, we have that
\begin{eqnarray*}
h_A(b_1b_2\ldots b_m ib_m\ldots b_2) & = & h_A(b_1b_2\ldots b_m)h_A(ib_m\ldots b_2)\\
& = & \mx{\alpha & *\\ 0 & \alpha\inv} \mx{\alpha\inv & *\\ 0 & \alpha}\\
& = & \mx{1 & *\\ 0 & 1}\in\cK ,
\end{eqnarray*}
and similarly for $h_A(ib_m\ldots b_2b_1b_2\ldots b_m)$.
\epf

\bcor\label{cor:yey}
Let $A=(A_0,A_1)$ be a pair of elements of $\SL{q}$, with
$$A_i=\mx{\xi_i & -1\\1 & 0}$$
for $i=0,1$.
Let $v=b_1b_2\ldots b_m$ be a bitstring such that $h_A(v)\in\cT$.
Let $u=b_m\ldots b_2$.
Then the palindromic bitstring $uv:=b_m\ldots b_1\ldots b_m$ of length $2m-1$ satisfies
$h_A(0uv1)=h_A(1uv0)$, a collision of length $2m+1$.
\ecor
\bpf
By Proposition \ref{prp:revcol}, we have that
\begin{eqnarray*}
h_A(v)h_A(0uv1)h_A(u) &= & h_A(v0u)h_A(v1u)\\
&=& h_A(v1u)h_A(v0u)\\
&=& h_A(v)h_A(1uv0)h_A(u).
\end{eqnarray*}
Multiplying on the right by $h_A(u)\inv$ and on the left by $h_A(v)$, the assertion follows.
\epf

We can now describe our algorithm. 
Let $A=(A_0,A_1)$ be a pair of generators of $SL_2(q)$ such that $|A_0 - A_1| = 0$.
First conjugate $A$ to matrices $B=(B_0,B_1)$ which have the form as in Lemma \ref{lem:rational}. 
As conjugation is a group isomorphism, the Cayley graph is unchanged, which thus remains an $\epsilon$-expander.
Note that the order of $\SL{q}$ is $(q-1)q(q+1)\approx q^3$.
By Proposition \ref{prp:exp}, we can generate bitstrings $v$ of length
$O(\log q/\epsilon^2)$ such that the statistical distance between $h_A(v)$
and a uniformly random element of $\SL{q}$ is smaller than $1/q^2$.


Next, hash with $h_B$ into the subgroup $\cT$ using a meet-in-the-middle approach as done by Petit et al.~\cite{PQTZ09}. We describe this approach using different, but equivalent terminology. 
In order to effectively hash into $\cT$, we need an efficient encoding of the cosets of $\cT$ in $\SL{q}$. 
In general, there is a bijective correspondence between transitive permutation groups and cosets of subgroups. 
The following proposition provides a concrete, efficient representation of these cosets as projective points.

\bdfn \label{dfn:projective}
Extend the definition of the quotient $\alpha\beta^{-1}$ to the case $\beta=0$ by declaring
$\alpha\cdot 0^{-1}=\infty$ for all $\alpha\in\bbF_q$.
\edfn

\bprp \label{Tcode}
The map
\begin{eqnarray*}
\SL{q}/\cT & \longrightarrow & \bbF_q\cup \{\infty\}\\
\mx{\alpha & \beta \\ \gamma & \delta}\cT & \longmapsto & \alpha\gamma^{-1}
\end{eqnarray*}
is well defined and bijective.
\eprp
\bpf
Assume that
$$\mx{\alpha_1 & * \\ \gamma_1 & *}\cT = \mx{\alpha_2 & * \\ \gamma_2 & *}\cT.$$
Then
$$
\mx{* & * \\ -\gamma_2 & \alpha_2}\mx{\alpha_1 & * \\ \gamma_1 & *} =
\mx{\alpha_2 & * \\ \gamma_2 & *}\inv\mx{\alpha_1 & * \\ \gamma_1 & *}\in\cT,$$
that is,
$-\gamma_2\alpha_1+\alpha_2\gamma_1=0$.
Thus, $\alpha_1\gamma_2=\alpha_2\gamma_1$, and the claim follows by considering the possible cases:
if any of $\gamma_1,\gamma_2$ is $0$, say,
$\gamma_1=0$, then $\alpha_1\neq 0$ (since the matrices are invertible),
and thus $\gamma_2=0$, and the code of both cosets is $\infty$.
If none of $\gamma_1,\gamma_2$ is $0$, then the codes are
$\alpha_1\gamma_1\inv = \alpha_2\gamma_2\inv$.
This proves that the map is well defined.

It is clear that the map is onto. As $|\SL{q}/\cT|=q+1=|\bbF_q\cup\{\infty\}|$, the map is bijective.
\epf

So, to hash into $\cT$, produce matrices $C$ by lazy random walks on the Cayley graph of
$(\SL{q}, \{g_0,g_1\})$, starting at $e$, together with bitstrings $v\in\{0,1\}^*$
of length $O(\log q)$ such that $C=h_B(v)$,
and store $v$ and the code of the coset $h_B(v)\cT$, as given by Proposition~\ref{Tcode}. That is, if $C=h_B(v)$
then in terms of the entries of $C$ the code of $C\cT$ is given by $c_{11}c_{21}\inv$.
Search for the code of $C\inv \cT$ in the set of stored codes.
The code of $C\inv\cT$, in terms of the entries of $C$, is $-c_{22}c_{21}\inv$.
If one is found, say of $h_B(u)$, then $h_B(u)\cT=h_B(v)\inv \cT$, and therefore $$h_B(vu)\cT=h_B(v)h_B(u)\cT=\cT,$$ so that we can terminate with
$$h_B(vu)\in \cT.$$

By Proposition \ref{prp:exp}, for each pair $u,v$ of our bitstrings,
the probability that the codes of $h_B(u)\cT$ and $h_B(v)\inv\cT$
are equal is, up to an additive $O(1/q^2)$ error, the same as the probability that the codes of $r_0\cT$ and $r_1\cT$
are equal, for uniformly random elements $r_0,r_1$ of $\SL{q}$. As $|\SL{q}/\cT|=q+1$, this probability is $1/(q+1)$.
The additive error of $O(1/q^2)$ is negligible compared to that, thus $O(\sqrt{q})$ bitstrings suffice for the above
procedure to terminate.

Suppose we have found a bitstring $b_1\ldots b_{2m}=uv$ whose hash value lies in $\cT$.
By Corollary \ref{cor:yey}, the palindromic bitstring
$$w:=b_{2m}\ldots b_1\ldots b_{2m}$$
satisfies
$$h_A(0w1)=h_A(1w0);$$
a collision of length $4m+1$, which is $O(\log q/\epsilon^2)$.
This completes the proof of Theorem \ref{thm:lincol}. \qed
\medskip

\brem\label{rem:hlin}
Heuristically, there is no need to assume in Theorem \ref{thm:lincol} that $A_0$ and $A_1$ generate $\SL{q}$.
Indeed, if they do not, then as shown in Lemma \ref{lem:rational}, they are simultaneously conjugate to elements of $\cT$,
and thus we can find a collision of length $\lg q$ as in Section \ref{subsec:second}. Thus, in any case we end up with
collisions of length roughly $\lg q$ if $|A_0-A_1|=0$.
\erem

\brem\label{rem:hashid}
Note that once a string $v$ is found that hashes into $\cK$ (as in Proposition~\ref{prp:revcol}) one can construct preimages to the identity element by concatenating $v$ with itself $p$ times.
\erem

\subsection*{Heuristic estimations and computer experiments.}

Throughout this paper, in our heuristic estimations we assume that for our purposes
hashes of distinct bitstrings
behave as if they are independent, uniformly distributed elements of the group in question.
(Unless there is an obvious obstruction, cf.\ Section \ref{sec:obs}.)

For the algorithm presented above, one needs that, for two of our generated matrices, $C_1,C_2$,
the codes of $C_1\cT$ and $C_2\inv\cT$ are identical.
This happens, heuristically, with probability $1/(q+1)\approx 1/q$.
Thus, we need to generate about $\sqrt{q}$ matrices. To this end, it suffices to hash all bitstrings of length up to
$\lg \sqrt{q}\approx\lg q/2$. Having achieved that, the length of the bitstring hashing to $\cT$ is twice that,
$\lg q$, and the length of the final collision is roughly $2\lg q$.
Our experimental results suggest that this heuristic is quite precisely correct.

We have tested our algorithms for a variety of pairs $p,n$ such that
$q=p^n\approx 2^{16},2^{32}$.
For each $N=16,32$, we first chose a random $p$ in a prescribed interval
$\{2^k,2^k+1,2^k+2,\dots,2^{k+1}\}$
indicated in the tables below, and then took $n$ to be
the rounded value of $N/\lg p$, so that $p^n\approx 2^N$.
For each choice of $N$ and an interval for $p$, we conducted $10{,}000$ experiments where, in each
experiment, we took a random $\xi_0,\xi_1\in\bbF_q$,
and applied our algorithm to the pair
$$A_0=\mx{\xi_0 & -1\\1 & 0},\;\; A_1=\mx{\xi_1 & -1\\1 & 0}.$$
The output of these sets of $10{,}000$ experiments is the minimum, median, average (and standard deviation), and maximum
values encountered for each of the measured quantities (work and length).
For $N=16$, we have also computed, for the same instances, the work needed to find the shortest collision (by breadth-first search enumeration) and its length.

\begin {table}[!h]
\begin{changemargin}{-5cm}{-5cm}
\caption{Results for $q\approx 2^{16}$:
Minimum, median, \textbf{average} (and standard deviation), and maximum
values encountered. $10{,}000$ experiments for each range of $p$.
}\label{tab:can16}

\begin{center}
\begin{tabular}{|l||l|l||l|l|}
\cline{2-5}
\multicolumn{1}{c||}{ } & \multicolumn{2}{|c||}{\bigstrut \textbf{shortest collision}} & \multicolumn{2}{|c|}{\textbf{our algorithm}}\\
\hline
\bigstrut $p\in$ & work & length & work & length\\
\hline
\hline

\bigstrut $\{2^{1},\dots,2^{2}\}$ & $0.00 q$ & $0.38\lg q$ & $0.04 \sqrt{q}$ & $0.70\lg q$ \\
 \bigstrut & $6.29 q$ & $1.12\lg q$ & $2.42\sqrt{q}$ & $2.20\lg q$ \\
 \bigstrut & $\mathbf{8.75}q$ ($9.41q$) & $\mathbf{1.10}\lg q$ ($0.12\lg q$) & $\mathbf{2.50}\sqrt{q}$ ($1.30\sqrt{q}$) & $\mathbf{2.16}\lg q$ ($0.24\lg q$) \\
\bigstrut & $100.90q$ & $1.38\lg q$ & $8.28\sqrt{q}$ & $2.68\lg q$ \\
 \hline
\bigstrut $\{2^{3},\dots,2^{4}\}$ & $0.00 q$ & $0.20\lg q$ & $0.01 \sqrt{q}$ & $0.40\lg q$ \\
 \bigstrut & $3.67 q$ & $1.08\lg q$ & $2.23\sqrt{q}$ & $2.22\lg q$ \\
 \bigstrut & $\mathbf{5.22}q$ ($5.14q$) & $\mathbf{1.05}\lg q$ ($0.12\lg q$) & $\mathbf{2.44}\sqrt{q}$ ($1.29\sqrt{q}$) & $\mathbf{2.16}\lg q$ ($0.26\lg q$) \\
\bigstrut & $61.65q$ & $1.35\lg q$ & $8.02\sqrt{q}$ & $2.78\lg q$ \\
 \hline
\bigstrut $\{2^{7},\dots,2^{8}\}$ & $0.00 q$ & $0.20\lg q$ & $0.01 \sqrt{q}$ & $0.32\lg q$ \\
 \bigstrut & $3.87 q$ & $1.08\lg q$ & $2.30\sqrt{q}$ & $2.20\lg q$ \\
 \bigstrut & $\mathbf{5.40}q$ ($5.33q$) & $\mathbf{1.06}\lg q$ ($0.12\lg q$) & $\mathbf{2.45}\sqrt{q}$ ($1.30\sqrt{q}$) & $\mathbf{2.16}\lg q$ ($0.26\lg q$) \\
\bigstrut & $56.84q$ & $1.34\lg q$ & $7.99\sqrt{q}$ & $2.84\lg q$ \\
 \hline
\bigstrut $\{2^{15},\dots,2^{16}\}$ & $0.00 q$ & $0.26\lg q$ & $0.02 \sqrt{q}$ & $0.46\lg q$ \\
 \bigstrut & $3.77 q$ & $1.07\lg q$ & $2.34\sqrt{q}$ & $2.20\lg q$ \\
 \bigstrut & $\mathbf{5.34}q$ ($5.29q$) & $\mathbf{1.05}\lg q$ ($0.12\lg q$) & $\mathbf{2.46}\sqrt{q}$ ($1.32\sqrt{q}$) & $\mathbf{2.16}\lg q$ ($0.26\lg q$) \\
\bigstrut & $69.47q$ & $1.35\lg q$ & $10.73\sqrt{q}$ & $2.74\lg q$ \\
 \hline

\end{tabular}
\end{center}
\end{changemargin}
\end{table}

\begin {table}[!h]
\caption{Results of the new algorithm for $q\approx 2^{32}$:
Minimum, median, \textbf{average} (and standard deviation), and maximum
values encountered. $10{,}000$ experiments for each range of $p$.}\label{tab:can32}
\begin{center}
\begin{tabular}{|l||l|l|}
\hline
\bigstrut $p\in$ & \textbf{work} & \textbf{length}\\
\hline
\hline

\bigstrut $\{2^{1},\dots,2^{2}\}$ & $0.04 \sqrt{q}$ & $1.34\lg q$ \\
 \bigstrut & $2.39\sqrt{q}$ & $2.10\lg q$ \\
 \bigstrut & $\mathbf{2.48}\sqrt{q}$ ($1.31\sqrt{q}$) & $\mathbf{2.08}\lg q$ ($0.12\lg q$) \\
\bigstrut & $9.03\sqrt{q}$ & $2.40\lg q$ \\
 \hline
\bigstrut $\{2^{7},\dots,2^{8}\}$ & $0.02 \sqrt{q}$ & $1.22\lg q$ \\
 \bigstrut & $2.33\sqrt{q}$ & $2.10\lg q$ \\
 \bigstrut & $\mathbf{2.48}\sqrt{q}$ ($1.30\sqrt{q}$) & $\mathbf{2.08}\lg q$ ($0.12\lg q$) \\
\bigstrut & $8.51\sqrt{q}$ & $2.40\lg q$ \\
 \hline
\bigstrut $\{2^{15},\dots,2^{16}\}$ & $0.03 \sqrt{q}$ & $1.28\lg q$ \\
 \bigstrut & $2.33\sqrt{q}$ & $2.10\lg q$ \\
 \bigstrut & $\mathbf{2.48}\sqrt{q}$ ($1.31\sqrt{q}$) & $\mathbf{2.08}\lg q$ ($0.12\lg q$) \\
\bigstrut & $8.27\sqrt{q}$ & $2.40\lg q$ \\
 \hline
\bigstrut $\{2^{31},\dots,2^{32}\}$ & $0.03 \sqrt{q}$ & $1.26\lg q$ \\
 \bigstrut & $2.37\sqrt{q}$ & $2.10\lg q$ \\
 \bigstrut & $\mathbf{2.48}\sqrt{q}$ ($1.30\sqrt{q}$) & $\mathbf{2.08}\lg q$ ($0.12\lg q$) \\
\bigstrut & $8.50\sqrt{q}$ & $2.36\lg q$ \\
 \hline

\end{tabular}
\end{center}
\end{table}

The results of our experiments are displayed in Tables \ref{tab:can16} and \ref{tab:can32}.
The striking observation is that, for all of these sets of parameters, and for the total $80{,}000$ experiments conducted,
none deviated substantially from our optimistic heuristic estimations.
Moreover, it is clearly visible that our algorithm is not sensitive to the field characteristic $p$.

\section{A generic short collision search algorithm}\label{sec:generic}

We now present a generic collision finding algorithm for $\SL{q}$ homomorphic hash functions for arbitrary $q$ and
arbitrary pairs $A=(A_0,A_1)$.
Heuristically, and according to experiments, our algorithm
finds collisions of length roughly $2\lg^2{q}/\lg\lg q$ in running time $O(\sqrt{q})$.
This algorithm improves upon an algorithm of Petit et al.~\cite{PQTZ09} for $q$ a power of $2$.
Petit et al.\ demonstrate, heuristically, that their algorithm
is expected to find collisions of length about $12{\lg}^2{q}$ in running time $O(\sqrt{q}\log{q})$.
A straightforward generalization of their algorithm to an arbitrary field size $q=p^n$ yields collisions of length about $12p{\lg}^2{q}$, and a slight modification of their approach yields $p$ times shorter collisions. We detail this approach and its
mentioned refinement in Appendix \ref{apx:PQTZ}.

The basic idea of our approach is to hash with $A=(A_0,A_1)$ until we find two elements that commute. For suppose we find two distinct strings $u,v$ whose hash values commute. Then a collision is given by $h_A(uv)=h_A(vu)$.
An obvious approach would be to hash into a commutative subgroup.

Roughly speaking, our algorithm is as follows.
The first step is to hash twice on $h_A$ into the subgroup $\cT$.
In fact, we show, heuristically, that we may assume that one of the matrices $A_0,A_1$ is already in $\cT$,
and it suffices to hash just once into $\cT$. This halves the amount of work, and makes it possible to
reduce the length of the final collision by a factor of $\lg\lg q$.
We then use the obtained matrices $C_0,C_1\in\cT$, to reduce the problem to hashing on $h_C$ to find two commuting elements. As we will see, aiming for the above-mentioned subgroup $\cK$ (this was the approach taken by Petit et al.~\cite{PQTZ09}) is problematic for our approach, whereas the subgroup $\cD$ of diagonal matrices is a good choice. In fact, we have a slightly better method, hashing directly to commuting elements, not necessarily diagonal ones.

We describe our algorithm in two phases: the first phase describes how to reduce the problem into one where $A_0,A_1$ are
in $\cT$, and the second phase describes how to hash on $\cT$ to find commuting elements.

\subsection{First phase: moving into $\cT$}\label{subsec:T}

In this phase we find two short bitstrings hashing into $\cT$. Finding the first string is easy. Since conjugation is a group automorphism, collisions are preserved under conjugation. The probability that a matrix in $\SL{q}$ is diagonalizable is $1/2-\Theta(1/q)$ \cite{SL2}. Thus heuristically, $A_0,A_1$ or short combination thereof, call it $A_2$, may be assumed to be diagonalizable. In other words, there is a bitstring $u_0$ of constant length such that $A_2:=h_A(u_0)$ is diagonalizable.

Let $P\in\SL{q}$ be such that $P\inv A_2 P$ is diagonal. In particular, $P\inv A_2 P\in \cT$. Conjugating $A_0,A_1$ by $P$, let
\begin{eqnarray*} B_0 & := & P\inv A_0 P,\\ B_1 & := & P\inv A_1 P. \end{eqnarray*} Setting $B=(B_0,B_1)$, we have that $$C_0:=h_B(u_0)=P\inv A_2 P\in \cT.$$

It remains to find a second string whose hash value on $h_B$ lies in $\cT$, which we can do using the meet-in-the-middle method used in the proof of Theorem~\ref{thm:lincol}.
Heuristically, we expect that among roughly $\sqrt{q}$ bitstrings there will be strings $u,v$
with the same code (as given by Proposition~\ref{Tcode}), so that the string $vu$ hashes into $\cT$ and
$$|vu|=|u|+|v|\approx 2 \lg \sqrt{q}=\lg q.$$
Setting $u_1:=vu$ we arrive at two strings $u_0,u_1$ of lengths $l_0$ constant and
$l_1\approx\lg q$, respectively, hashing to $C_0,C_1\in \cT$.

\subsection{Second phase: finding commuting elements in $\cT$}\label{subsec:second}

After finding strings $u_0,u_1$ hashing to $C_0,C_1\in \cT$, the next and final step is to find two strings whose hash values commute on $h_C$.

\subsubsection{An obstruction}\label{sec:obs}

It is tempting to repeat the same procedure for $h_C$ and the subgroup $\cK$ of $\cT$ of index $q-1$. Unfortunately, we encounter the following obstruction, stemming from $\cK$ being abelian. Let
$$T_0=\mx{\alpha_0 & *\\ 0 & \alpha_0\inv},\;\; T_1=\mx{\alpha_1 & *\\ 0 & \alpha_1\inv}.$$
For each bitstring $b_1\ldots b_m$, the upper left entry of $T_{b_1}\ldots T_{b_m}$ is $$\alpha_{b_0}\ldots\alpha_{b_m}=\alpha_0^{\nu_0(b_1\ldots b_m)}\alpha_1^{\nu_1(b_1\ldots b_m)},$$
where $\nu_0(\cdot),\nu_1(\cdot)$ denote, respectively, the number of $0$-bits and the number of $1$-bits in a bitstring.

On average, to have $\alpha_0^{k_0}\alpha_1^{k_1}=1$, we need $k_0$ and $k_1$ to be roughly $\sqrt{q}$, which would increase the length of the final collision by $\sqrt{q}$, i.e. exponentially in $\lg q$.

This problem is circumvented by Petit et al.~\cite{PQTZ09} by hashing roughly $\lg q$ times into $\cT$, and then using an algorithm based on the LLL algorithm and computing discrete logarithms in $\bbF_q$ (see Appendix \ref{apx:PQTZ}). However, this has a price, both in terms of running time and the length of resulting collisions.

We propose two simpler and more efficient approaches.

\subsubsection{First solution: hashing into $\cD$}

Instead of hashing into $\cK$, consider the subgroup $\cD$ of $\cT$, consisting of the diagonal matrices
$$\mx{\alpha & 0\\ 0 & \alpha\inv}$$
for nonzero $\alpha\in\bbF_q$.
To construct a collision, we need to find two strings that hash on $h_C$ into $\cD$. We already have one such string, namely $u_0$ with hash value $h_B(u_0):=h_C(0)$.

We can employ a similar meet-in-the-middle approach as in the previous phase to find a bitstring $w$ of length roughly $\lg q$ such that $h_C(w)\in\cD$. Note that to avoid trivialities $w$ must not be a sequence of concatenations of $u_0$.

Again, to employ a meet-in-the-middle approach we need an efficient encoding of the cosets of $\cD$ in $\cT$, which is given by the following.
\bprp
The map
\begin{eqnarray*}
\cT/\cD & \longrightarrow & \bbF_q\\
\mx{\alpha & \beta \\ 0 & \alpha\inv}\cD & \longmapsto & \alpha\beta
\end{eqnarray*}
is well defined and bijective.
\eprp
\bpf
Assume that
$$\mx{\alpha_1 & \beta_1 \\ 0 & \alpha_1\inv}\cD = \mx{\alpha_2 & \beta_2 \\ 0 & \alpha_2\inv}\cD.$$
Then
$$
\mx{\alpha_2\inv & -\beta_2 \\ 0 & \alpha_2}\mx{\alpha_1 & \beta_1 \\ 0 & \alpha_1\inv} =
\mx{\alpha_2 & \beta_2 \\ 0 & \alpha_2\inv}\inv\mx{\alpha_1 & \beta_1 \\ 0 & \alpha_1\inv}\in\cD,$$
and therefore $\alpha_2\inv\beta_1-\beta_2\alpha_1\inv=0$, that is,
$\alpha_1\beta_1=\alpha_2\beta_2$, and the codes are equal.

The map is onto. As $|\cT/\cD|=q(q-1)/(q-1)=q=|\bbF_q|$, the map is bijective.
\epf

\subsubsection{Second solution: hashing to commuting elements of $\cT$}
This solution, which seeks for more balanced strings whose hashes commute,
turns out slightly better than the previous approach of hashing into $\cD$. We need a code to test when two elements of $\cT$ commute.

\bprp
Matrices
$$\mx{\alpha & \beta\\ 0 & \alpha\inv},\mx{\gamma & \delta\\ 0 & \gamma\inv}$$
not equal to $\pm I$ commute if and only if $(\alpha-\alpha\inv)\beta\inv=(\gamma-\gamma\inv)\delta\inv$.
\eprp
\bpf
By direct calculation, all entries of
$$
\mx{\alpha & \beta\\ 0 & \alpha\inv}\mx{\gamma & \delta\\ 0 & \gamma\inv}-
\mx{\gamma & \delta\\ 0 & \gamma\inv}\mx{\alpha & \beta\\ 0 & \alpha\inv}
$$
are $0$, except perhaps the upper right one
$$\alpha \delta + \beta \gamma\inv-\gamma\beta-\delta\alpha\inv =
\delta(\alpha-\alpha\inv)-\beta(\gamma-\gamma\inv),$$
which is $0$ if and only if
$$\delta(\alpha-\alpha\inv)=\beta(\gamma-\gamma\inv).$$

If $\beta$ and $\delta$ are both nonzero then we can rewrite the above equation as
$$(\alpha-\alpha\inv)\beta\inv=(\gamma-\gamma\inv)\delta\inv,$$ and the claim is proved.

If $\beta=0$ then, since $\alpha \neq \pm 1$ we have that $\delta(\alpha-\alpha\inv)=0$ implies $\delta=0$. It follows that the matrices are diagonal, and thus commute, and we have that (in the notation of Definition~\ref{dfn:projective}) $$(\alpha-\alpha\inv)\beta\inv=\infty=(\gamma-\gamma\inv)\delta\inv.$$ The case $\delta=0$ is identical.
\epf

Thus, to find two strings whose hashes on $h_C$ commute do the following. For roughly $\sqrt{q}$ bitstrings $v$ (that are not a power of $u_0$) compute $$h_C(v) = \mx{\alpha & \beta\\ 0 & \alpha\inv}$$ and store $v$ and the code $(\alpha-\alpha\inv)\beta\inv$. If we ever encounter the code $0$ or $\infty$ then we are done, since this matrix commutes with $C_0$. Assuming this rare event does not occur, find two strings $u,v$ such that the codes of $h_C(u)$ and $h_C(v)$ are equal. We expect $$|uv|,|vu|\approx 2|v|\approx 2\lg\sqrt{q}=\lg q,$$ and  the overall length of the collision $$h_C(uv)=h_C(u)h_C(v)=h_C(v)h_C(u)=h_C(vu)$$ is on average, in terms of the original hash function $h_A$, $$|uv|\approx\frac{l_0+\lg q}{2}\cdot\lg q\approx\frac{1}{2}{\lg}^2 q.$$ The factor $1/2$ comes from expecting a roughly equal number of zeros and ones.

\subsection{Compressed collisions}
In the first phase, we arrived at two strings $u_0,u_1$ of lengths $l_0$ constant and
$l_1\approx\lg q$, respectively, hashing to $C_0,C_1\in \cT$.
For both the first and second solutions above, we can reduce the total collision length by exploiting the fact that $u_0$ is roughly $\lg q/l_0$ times shorter than $u_1$.

Let $C=(C_0,C_1)$.
For each bitstring $b_1\dots b_k\in\V$,
$$h_C(b_1\dots b_k)=C_{b_1}\cdots C_{b_k}=h_A(u_{b_1})\cdots h_A(u_{b_k})=h_A(u_{b_1}\dots u_{b_k}).$$
Define
\begin{eqnarray*}
\nu_0(b_1\dots b_k) & = & \left|\set{i=1,\dots,k}{b_i=0}\right|\\
\nu_1(b_1\dots b_k) & = & \left|\set{i=1,\dots,k}{b_i=1}\right|\\
\|b_1\dots b_k\|_{l_0,l_1}    & = & \nu_0(b_1\dots b_k)\cdot l_0+\nu_1(b_1\dots b_k)\cdot l_1.
\end{eqnarray*}
Then, in terms of $h_A$, the length of a collision $h_C(w_1)=h_C(w_2)$ is
$\max\{\|w_1\|_{l_0,l_1},\|w_2\|_{l_0,l_1}\}$.

Following is an algorithm for producing finite bitstrings $v$ such that the length $\|v\|_{l_0,l_1}$ is monotonically increasing,
for $l_0<l_1$.

\balg\label{alg:fib}
\mbox{}
\be
\itm $g := \gcd(l_0,l_1); \; k_0 := l_0/g; \; k_1 := l_1/g$.
\itm For $n=1,\dots,k_1$:
$$S_n:=
\begin{cases}
\{0^{n/k_0}\} & \mbox{if } k_0\mid n,\\
\emptyset & \mbox{otherwise.}
\end{cases}
$$
\itm $S_{k_1} := S_{k_1}\cup \{1\}$.
\itm For $n=k_1+1,k_1+2,\dots$:
$$S_n := \set{v0}{v \in S_{n-k_0}} \cup \set{v1}{v \in S_{n-k_1}}.$$
\ee
\ealg

\bprp\label{prp:fib}
Let $l_0<l_1$ be natural numbers. In the notation of Algorithm \ref{alg:fib}:
\be
\itm For each $v\in \V$, $\|v\|_{l_0,l_1}$ is divisible by $g$.
\itm For each $n$,  $S_n = \set{v\in \V}{\|v\|_{l_0,l_1}/g=n}$.
\itm $|S_n|=|S_{n-k_0}|+|S_{n-k_1}|$, a generalized Fibonacci sequence.
\itm $|S_1\cup S_2\cup \ldots\cup S_n|\ge \lfloor k_1/k_0\rfloor^{\lfloor n/2k_1\rfloor}=\lfloor l_1/l_0\rfloor^{\lfloor gn/2l_1\rfloor}$.
\ee
\eprp
\bpf
(1) Obvious.

(2,3) By induction on $n$, observing that the bitstrings of length $gn$ split into those terminating with $0$ and
those terminating with $1$.

(4) Let $m=\lfloor n/2k_1\rfloor$.
The map
\begin{eqnarray*}
\{1,\dots,\lfloor k_1/k_0\rfloor\}^m & \longrightarrow & S_1\cup S_2\cup \ldots\cup S_n\\
(i_1,\dots,i_m) & \longmapsto & 0^{i_1}10^{i_2}1\ldots 0^{i_m}1
\end{eqnarray*}
is injective.
Its range is as claimed. Indeed,
$$\|0^{i_1}10^{i_2}1\ldots 0^{i_m}1\|_{l_0,l_1}\le m(l_0\lfloor k_1/k_0\rfloor+l_1)=
m(l_0\lfloor l_1/l_0\rfloor+l_1)\le m\cdot 2l_1\le (n/2k_1)\cdot 2l_1 = gn.$$
Apply (2).
\epf

To find shorter collisions we use the same algorithms as before, but generate the bitstrings according to Algorithm \ref{alg:fib}.
By item (4) of Proposition \ref{prp:fib},
we need that
$$\sqrt{q}\approx(l_1/l_0)^{gn/2l_1},$$
and since $l_0$ is constant, we have
$$\frac{1}{2}\lg q\approx \frac{gn}{2l_1}\lg\frac{l_1}{l_0}\approx
\frac{gn}{2}\frac{\lg{l_1}}{l_1},$$
that is,
$$gn\approx \frac{l_1\lg q}{\lg l_1}\approx \frac{{\lg}^2 q}{\lg\lg q}.$$
The length of the obtained collision is twice that.

\brem
The diagonalization trick in the first phase, that reduces the running time by a constant factor,
leads to the $\lg\lg q$ factor reduction of the resulting length. It may be that the constant
estimation for the minimal length of a diagonal element is not provable,
even using that the Cayley graph of $(\SL{q},\{A_0,A_1\})$ is an expander. The reason is that
a random walk in an expander graph may miss a subset of half the size of the group
for a logarithmic number of steps.
If we aim, instead, at collisions of length $O(\log^2 q)$,
then the first phase of our algorithm would be to hash twice into $\cT$, and the estimations for running time and
bitstring lengths are  provable as in the previous section.
We do not know whether estimations in the second (noncompressed) phase are provable.
If, for two random elements $A_0,A_1$ of $\cT$, the Cayley graph of
$(\cT,\{A_0^{\pm 1},A_1^{\pm 1}\})$ is (with high probability) an expander, then they are.
\erem


\subsection{Computer experiments}
Computer experiments are reported in Tables \ref{tab:gen16} and \ref{tab:gen32}.
Here too, our optimistic estimations are all validated. Indeed, our estimation $2\lg^2 q/\lg\lg q$ turns out
slightly more generous than needed.


\begin{sidewaystable}
\centering
\caption{Results for $q\approx 2^{16}$:
Minimum, median, \textbf{average} (and standard deviation), and maximum
values encountered.
10{,}000 experiments for each range of $p$.
$L:=\lg^2 q/\lg\lg q$.}\label{tab:gen16}
\hspace*{0cm}
\vspace*{-16cm}
{\scriptsize
\begin{tabular}{|l||l|l||l|l|l|l|l|}
\cline{2-8}
\multicolumn{1}{c||}{ } &
\multicolumn{2}{|c||}{\bigstrut \textbf{shortest collision}} &
\textbf{diagonalizable} &
\multicolumn{2}{|c|}{\textbf{shortest triangular}} &
\multicolumn{2}{|c|}{\textbf{compressed search}}\\
\hline
\bigstrut $p\in$ & work & length & length & work & length & work & length\\
\hline
\hline

\bigstrut $\{2^{1},\dots,2^{2}\}$ & $0.00 q$ & $0.31\lg q$ & $1.00$ & $0.03 \sqrt{q}$ & $0.31\lg q$ & $0.00 \sqrt{q}$ & $0.08 L$ \\
 \bigstrut & $3.66 q$ & $1.07\lg q$ & $1.00$ & $2.41 \sqrt{q}$ & $1.06\lg q$ & $2.68 \sqrt{q}$ & $1.34 L$ \\
 \bigstrut & $\mathbf{5.38}q$ ($5.32q$) & $\mathbf{1.05}\lg q$ ($0.11\lg q$) & $\mathbf{1.40}$ ($0.79$) & $\mathbf{2.66}\sqrt{q}$ ($1.51\sqrt{q}$) & $\mathbf{1.04}L$ ($0.13\lg q$) & $\mathbf{2.89}\sqrt{q}$ ($1.54\sqrt{q}$) & $\mathbf{1.36}L$ ($0.30 L$) \\
\bigstrut & $54.98 q$ & $1.32\lg q$ & $5.00$ & $10.79 \sqrt{q}$ & $1.32\lg q$ & $12.13 \sqrt{q}$ & $2.98 L$ \\
 \hline
\bigstrut $\{2^{3},\dots,2^{4}\}$ & $0.00 q$ & $0.27\lg q$ & $1.00$ & $0.03 \sqrt{q}$ & $0.20\lg q$ & $0.00 \sqrt{q}$ & $0.08 L$ \\
 \bigstrut & $3.71 q$ & $1.08\lg q$ & $1.00$ & $2.45 \sqrt{q}$ & $1.04\lg q$ & $2.60 \sqrt{q}$ & $1.32 L$ \\
 \bigstrut & $\mathbf{5.35}q$ ($5.34q$) & $\mathbf{1.05}\lg q$ ($0.11\lg q$) & $\mathbf{1.41}$ ($0.79$) & $\mathbf{2.68}\sqrt{q}$ ($1.54\sqrt{q}$) & $\mathbf{1.04}L$ ($0.13\lg q$) & $\mathbf{2.81}\sqrt{q}$ ($1.51\sqrt{q}$) & $\mathbf{1.34}L$ ($0.32 L$) \\
\bigstrut & $58.57 q$ & $1.35\lg q$ & $6.00$ & $10.34 \sqrt{q}$ & $1.35\lg q$ & $11.56 \sqrt{q}$ & $3.02 L$ \\
 \hline
\bigstrut $\{2^{7},\dots,2^{8}\}$ & $0.00 q$ & $0.20\lg q$ & $1.00$ & $0.03 \sqrt{q}$ & $0.21\lg q$ & $0.00 \sqrt{q}$ & $0.06 L$ \\
 \bigstrut & $3.96 q$ & $1.08\lg q$ & $1.00$ & $2.40 \sqrt{q}$ & $1.06\lg q$ & $2.64 \sqrt{q}$ & $1.34 L$ \\
 \bigstrut & $\mathbf{5.54}q$ ($5.41q$) & $\mathbf{1.06}\lg q$ ($0.12\lg q$) & $\mathbf{1.41}$ ($0.79$) & $\mathbf{2.67}\sqrt{q}$ ($1.57\sqrt{q}$) & $\mathbf{1.04}L$ ($0.14\lg q$) & $\mathbf{2.81}\sqrt{q}$ ($1.48\sqrt{q}$) & $\mathbf{1.34}L$ ($0.32 L$) \\
\bigstrut & $49.42 q$ & $1.34\lg q$ & $5.00$ & $33.43 \sqrt{q}$ & $1.58\lg q$ & $9.88 \sqrt{q}$ & $2.86 L$ \\
 \hline
\bigstrut $\{2^{15},\dots,2^{16}\}$ & $0.00 q$ & $0.26\lg q$ & $1.00$ & $0.01 \sqrt{q}$ & $0.19\lg q$ & $0.00 \sqrt{q}$ & $0.06 L$ \\
 \bigstrut & $3.82 q$ & $1.07\lg q$ & $1.00$ & $2.43 \sqrt{q}$ & $1.06\lg q$ & $2.62 \sqrt{q}$ & $1.32 L$ \\
 \bigstrut & $\mathbf{5.41}q$ ($5.32q$) & $\mathbf{1.06}\lg q$ ($0.11\lg q$) & $\mathbf{1.41}$ ($0.80$) & $\mathbf{2.66}\sqrt{q}$ ($1.51\sqrt{q}$) & $\mathbf{1.04}L$ ($0.13\lg q$) & $\mathbf{2.82}\sqrt{q}$ ($1.52\sqrt{q}$) & $\mathbf{1.34}L$ ($0.32 L$) \\
\bigstrut & $51.93 q$ & $1.32\lg q$ & $5.00$ & $10.13 \sqrt{q}$ & $1.32\lg q$ & $10.52 \sqrt{q}$ & $2.88 L$ \\
 \hline

\end{tabular}
}
\end{sidewaystable}

\begin{sidewaystable}
\centering
\caption{Results for $q\approx 2^{32}$:
Minimum, median, \textbf{average} (and standard deviation), and maximum
values encountered.
10{,}000 experiments for each range of $p$.
$L:=\lg^2 q/\lg\lg q$.}\label{tab:gen32}
\hspace*{0cm}
\vspace*{-16cm}
\begin{tabular}{|l||l|l||l|l|l|}
\cline{2-6}
\multicolumn{1}{c||}{ } &
\bigstrut \textbf{diagonalizable} &
\multicolumn{2}{|c|}{\textbf{shortest triangular}} &
\multicolumn{2}{|c|}{\textbf{compressed search}}\\
\hline
\bigstrut $p\in$ & length & work & length & work & length\\
\hline
\hline

\bigstrut $\{2^{1},\dots,2^{2}\}$ & $1.00$ & $0.04 \sqrt{q}$ & $0.66\lg q$ & $0.03 \sqrt{q}$ & $0.58 L$ \\
 \bigstrut & $1.00$ & $2.44 \sqrt{q}$ & $1.03\lg q$ & $2.98 \sqrt{q}$ & $1.34 L$ \\
 \bigstrut & $\mathbf{1.41}$ ($0.80$) & $\mathbf{2.69}\sqrt{q}$ ($1.54\sqrt{q}$) & $\mathbf{1.02}\lg q$ ($0.06\lg q$) & $\mathbf{3.23}\sqrt{q}$ ($1.78\sqrt{q}$) & $\mathbf{1.38}L$ ($0.22 L$) \\
\bigstrut & $6.00$ & $10.38 \sqrt{q}$ & $1.17\lg q$ & $10.66 \sqrt{q}$ & $2.44 L$ \\
 \hline
\bigstrut $\{2^{7},\dots,2^{8}\}$ & $1.00$ & $0.02 \sqrt{q}$ & $0.56\lg q$ & $0.00 \sqrt{q}$ & $0.18 L$ \\
 \bigstrut & $1.00$ & $2.39 \sqrt{q}$ & $1.02\lg q$ & $2.92 \sqrt{q}$ & $1.34 L$ \\
 \bigstrut & $\mathbf{1.41}$ ($0.81$) & $\mathbf{2.63}\sqrt{q}$ ($1.51\sqrt{q}$) & $\mathbf{1.02}\lg q$ ($0.07\lg q$) & $\mathbf{3.19}\sqrt{q}$ ($1.74\sqrt{q}$) & $\mathbf{1.38}L$ ($0.22 L$) \\
\bigstrut & $5.00$ & $10.07 \sqrt{q}$ & $1.21\lg q$ & $11.19 \sqrt{q}$ & $2.42 L$ \\
 \hline
\bigstrut $\{2^{15},\dots,2^{16}\}$ & $1.00$ & $0.04 \sqrt{q}$ & $0.65\lg q$ & $0.00 \sqrt{q}$ & $0.18 L$ \\
 \bigstrut & $1.00$ & $2.43 \sqrt{q}$ & $1.03\lg q$ & $2.92 \sqrt{q}$ & $1.34 L$ \\
 \bigstrut & $\mathbf{1.40}$ ($0.79$) & $\mathbf{2.69}\sqrt{q}$ ($1.55\sqrt{q}$) & $\mathbf{1.02}\lg q$ ($0.07\lg q$) & $\mathbf{3.19}\sqrt{q}$ ($1.77\sqrt{q}$) & $\mathbf{1.38}L$ ($0.22 L$) \\
\bigstrut & $5.00$ & $10.26 \sqrt{q}$ & $1.19\lg q$ & $10.52 \sqrt{q}$ & $2.46 L$ \\
 \hline
\bigstrut $\{2^{31},\dots,2^{32}\}$ & $1.00$ & $0.01 \sqrt{q}$ & $0.51\lg q$ & $0.00 \sqrt{q}$ & $0.14 L$ \\
 \bigstrut & $1.00$ & $2.43 \sqrt{q}$ & $1.03\lg q$ & $2.95 \sqrt{q}$ & $1.34 L$ \\
 \bigstrut & $\mathbf{1.42}$ ($0.80$) & $\mathbf{2.67}\sqrt{q}$ ($1.54\sqrt{q}$) & $\mathbf{1.02}\lg q$ ($0.06\lg q$) & $\mathbf{3.21}\sqrt{q}$ ($1.75\sqrt{q}$) & $\mathbf{1.38}L$ ($0.22 L$) \\
\bigstrut & $6.00$ & $10.61 \sqrt{q}$ & $1.17\lg q$ & $11.17 \sqrt{q}$ & $2.40 L$ \\
 \hline

\end{tabular}
\end{sidewaystable}


\section{Linear collisions for $q=2^n$}\label{sec:even}

Faug\`{e}re et al.\ \cite{Faugere}, building on \cite{PetitTowards}, devised a heuristic subexponential time
algorithm in the case where $q$ is a power of $2$.
Heuristically, for $n_0\le n$, their time complexity and collision length are
$$2^{\mbox{\normalsize$\frac{\omega n\log n\log n_0}{n_0\log(n/n_0)}$}} \mbox{ and } \frac{32n^33^{n_0}}{n_0},$$
respectively, where $\omega\approx 2.8$ is the matrix multiplication constant.
For the collisions to have polynomial length, $n_0$ must be $O(\log n)$.
To minimize time complexity, $n_0$ should be $\Theta(\log n)$.
Let $n_0=c\log n$. Then the time complexity and collision length are, very roughly,
$$2^{\mbox{\normalsize$\frac{\omega}{c}\cdot\frac{n\log\log n}{\log n}$}} \mbox{ and }
\frac{32}{c}\cdot \frac{n^{3+c\log 3}}{\log n}.$$
To compare the performance of our algorithm to that of the subexponential algorithm from a practical point of view,
we have limited the length of the collision to $2^{80}$ bits (one terra terra bits), a generous upper bound for
an acceptable message length. Then, for each $n=64,128,256,\dots,16384$, we have computed the maximal
value of $n_0$ for which the collision length of the subexponential algorithm is not greater than $2^{80}$.
For this value of $n_0$, the running time of the subexponential algorithm is minimal.
Table \ref{tab:compar1} lists, for each of these $n$, the running time and collision length (rounded) for our algorithm
and the subexponential one. One sees clearly that, limiting the collision length to $2^{80}$,
our generic algorithm is much faster in all cases, and produces much shorter collisions.

\begin{table}
\caption{Generic collision search versus subexponential collision search.}\label{tab:compar1}
\begin{tabular}{|c||c|c||c|c|}
\cline{2-5}
\multicolumn{1}{c||}{ } & \multicolumn{2}{|c||}{\textbf{subexponential algorithm}} & \multicolumn{2}{|c|}{\bigstrut \textbf{our algorithm}}\\
\hline
\bigstrut $q$ & work & length & work & length\\
\hline
\hline
\bigstrut $2^{64}$ & $2^{143}$ & $2^{80}$ & $2^{32}$ & $2^{10}$\\
\bigstrut $2^{128}$ & $2^{137}$ & $2^{80}$ & $2^{64}$ & $2^{12}$\\
\bigstrut $2^{256}$ & $2^{202}$ & $2^{80}$ & $2^{128}$ & $2^{14}$\\
\bigstrut $2^{512}$ & $2^{344}$ & $2^{80}$ & $2^{256}$ & $2^{16}$\\
\bigstrut $2^{1024}$ & $2^{625}$ & $2^{80}$ & $2^{512}$ & $2^{18}$\\
\bigstrut $2^{2048}$ & $2^{1181}$ & $2^{80}$ & $2^{1024}$ & $2^{20}$\\
\bigstrut $2^{4096}$ & $2^{2292}$ & $2^{80}$ & $2^{2048}$ & $2^{21}$\\
\bigstrut $2^{8192}$ & $2^{4532}$ & $2^{80}$ & $2^{4096}$ & $2^{23}$\\
\bigstrut $2^{16384}$ & $2^{9093}$ & $2^{80}$ & $2^{8192}$ & $2^{25}$\\
\hline
\end{tabular}
\end{table}

But this is not the end of the story. Petit has realized that, for $q$ a power of $2$, given
the methods of \cite{PetitTowards} and \cite{Faugere}, our methods from Section
\ref{sec:rational} imply that, heuristically, collisions of linear length can be found
for \emph{arbitrary} generators $A_0,A_1$ of $\SL{2^n}$.
Modulo our results, this algorithm is implicit in the proof of Proposition 3 of \cite{Faugere}.
Following is a detailed description of this algorithm.

A matrix $E\in\SL{q}$ is \emph{orthogonal} if $E\transpose{E}=I$. The orthogonal matrices in
$\SL{2^n}$ are precisely matrices of the form
$$E=\mx{\alpha+1 & \alpha\\ \alpha & \alpha+1},$$
where $\alpha\in\bbF_{2^n}$ \cite{PetitTowards}. In particular, these matrices are symmetric and satisfy
$E^2=I$.

Let $A_0,A_1$ be generators of $\SL{2^n}$.
Let
$$B_0:=A_0A_1;\; B_1:=A_1A_0.$$
It suffices to find a collision for $(B_0,B_1)$.
The traces of $B_0$ and $B_1$ are equal. By the proof of
\cite[Lemma 2]{PetitTowards}, there are several possibilities:
\be
\item Certain (rare) pathologies happen,\footnote{Rare pathologies
are possible if $A_0,A_1$ are chosen in very special form, see the proof of \cite[Lemma 2]{PetitTowards}.}
in which there are collisions of length $2$, and we are done.
\item $B_0,B_1$ are simultaneously conjugate to upper triangular matrices, so by Section \ref{subsec:second}
we can find a collision of length $\lg q$ in time $\sqrt{q}$. This case is also rare for random generators.
\item In the remaining, main case,
$B_0,B_1$ can be simultaneously conjugated to a pair of the form
$$C,\transpose{C},$$
i.e., such that the second matrix is the transpose of the first. It suffices to find
a collision for $(C,\transpose{C})$. This is the only case remaining to be dealt with.
\ee
By \cite[Lemma 8]{PetitTowards}, we can find an \emph{orthogonal} matrix $E$ such that
$ECE=\transpose{C}$. Thus,
\begin{eqnarray*}
CE & = & E\transpose{C}, \mbox{ and}\\
\transpose{C}E & = & EC.
\end{eqnarray*}
Consider the pair $(CE,C)$.
By the above-mentioned special form of orthogonal matrices, $|E-I|=0$, and thus
$$|CE-C|=|C(E-I)|=|C|\cdot |E-I|=0.$$
Transforming a collision for $(CE,C)$ to one for $(C,C^t)$ is possible if the number of $0$'s
is either even in both strings or odd in both strings. In this case, using that
$CE = E\transpose{C}$, $\transpose{C}E = EC$, and $E^2=I$,
the $E$'s can be pushed to the left,
transposing the matrices $C,\transpose{C}$ on their way, and vanishing when meeting other $E$'s.
If an $E$ remains (necessarily, on both sides), it can be canceled from both sides.
Thus, heuristically, we need two collisions for $(CE,C)$ to conclude.

By the proof of Lemma \ref{lem:rational}, there are two cases to consider:
$CE$ and $C$ are simultaneously conjugate to either upper triangular matrices
or to matrices of the form
$$\mx{\xi_i & 1\\ 1 & 0}.$$
In the former case,
by Section \ref{subsec:second}, we can find collisions of the prescribed form of length $\lg q$ in time $\sqrt{q}$.
In the latter, main case, by Corollary \ref{cor:yey} it suffices to hash with $h=h_{(CE,C)}$ once
(in time $\sqrt{q}$ and string length $\lg q$) into an upper triangular matrix, say $h(b_1\dots b_m)\in\cT$.
By Corollary \ref{cor:yey},
$$h(0b_m\dots b_1\dots b_m1)=h(1b_m\dots b_1\dots b_m0).$$
As the number of $0$ bits in both strings of this collision is equal, this
collision can be transformed into one for $(C,\transpose{C})$, and we are done.

To illustrate this algorithm in the main case, assume that $A_0,A_1$ are given. Then:
\begin{enumerate}
\item Set $B_0:=A_0A_1,\; B_1:=A_1A_0.$
\item Find a matrix $P$ such that $P\inv B_0 P=C, \; P\inv B_1P=\transpose{C},$ for some matrix $C$.
\item Find an orthogonal matrix $E$ such that $CE = E\transpose{C}$.
\item Find a matrix $Q$ such that $$D_0:=Q\inv (CE) Q=\mx{\xi_0 & 1\\1 & 0},\;\; D_1:=Q\inv CQ=\mx{\xi_1 & 1\\1 & 0}.$$
\item Find a bitstring $b_1\dots b_m$ such that $D_{b_1}\cdots D_{b_m}\in\cT$, so that
$$D_0D_{b_m}\cdots D_{b_1}\cdots D_{b_m}D_1 = D_1D_{b_m}\cdots D_{b_1}\cdots D_{b_m}D_0.$$
\end{enumerate}
For example, assume that $m=3$ and $b_1\dots b_m=011$. Then
$$D_0 D_0D_1D_1D_1D_0 D_1 = D_1 D_0D_1D_1D_1D_0 D_0,$$
and in terms of $CE$ and $C$,
$$CE  CE C C C CE  C  = C  CE C C C CE  CE.$$
Moving the $E$'s to the left, using $CE = E\transpose{C}$, $\transpose{C}E = EC$, and $E^2=I$,
we have that
$$E\transpose{C} C \transpose{C} \transpose{C} \transpose{C} \transpose{C}  C  =
E\transpose{C}  \transpose{C} C C C C \transpose{C},$$
and thus
$$\transpose{C} C \transpose{C} \transpose{C} \transpose{C} \transpose{C}  C  =
\transpose{C}  \transpose{C} C C C C \transpose{C}.$$
In terms of $B_0$ and $B_1$, we have that
$$B_1 B_0 B_1 B_1 B_1 B_1  B_0  =
B_1  B_1 B_0 B_0 B_0 B_0 B_1,$$
and in terms of $A_0$ and $A_1$,
$$A_1A_0 A_0A_1 A_1A_0 A_1A_0 A_1A_0 A_1A_0  A_0A_1  =
A_1A_0  A_1A_0 A_0A_1 A_0A_1 A_0A_1 A_0A_1 A_1A_0.$$

The first reduction doubles the collision length. All other reductions preserve the collision length.
Thus, we expect collision lengths of the algorithm to be roughly
$$2\cdot 2\lg q=4\lg q.$$

\subsection{Computer experiments}
The results for $q=2^{16},2^{32}$ are very similar to those in Tables \ref{tab:can16} and \ref{tab:can32},
with the only difference that, as expected, the collision length is doubled.
Results of experiments for $q=2^{40}$ are provided in Table \ref{tab:even40}.
Here too, our heuristic estimations are confirmed, and even generous. The standard deviation of the collision
length is very small, and is expected to converge to $0$ as $q$ increases.

\begin {table}[!h]
\caption{Results of the new algorithm for $q=2^{40}$,
$10{,}000$ experiments.}\label{tab:even40}
\begin{center}
\begin{tabular}{|l||l|l|}
\hline
\bigstrut $q=2^{40}$ & \textbf{work} & \textbf{length}\\
\hline
\hline
\bigstrut Minimum & $0.02 \sqrt{q}$ & $2.76\lg q$ \\
\bigstrut Median & $2.41\sqrt{q}$ & $4.16\lg q$ \\
\bigstrut \textbf{Average} (and standard deviation) & $\mathbf{2.49}\sqrt{q}$ ($1.31\sqrt{q}$) & $\mathbf{4.14}\lg q$ ($0.2\lg q$) \\
\bigstrut Maximum & $8.32\sqrt{q}$ & $4.54\lg q$ \\
 \hline
\end{tabular}
\end{center}
\end{table}

\subsection*{Acknowledgments}
We thank Alexei Belov, Alex Lubotzky and Terrence Tao for useful information about expander Cayley graphs.
We also thank Emmanuel Breuillard for pointing out Proposition \ref{prp:exp}(1) to us, and for
his permission to include it here.
We owe special thanks to Christophe Petit, for numerous discussions about the problem studied here and the known methods,
for useful advice that helped improve the presentation of this paper, and for his observation that,
given the methods of \cite{PetitTowards} and \cite{Faugere}, our methods imply that, heuristically, collisions of linear length can be found for arbitrary generators of $\SL{2^n}$.
This research was initiated when the second named author visited
Simon Blackburn, Carlos Cid, and the first named author at Royal Holloway, University of London.
This author thanks his hosts for their kind hospitality.

\appendix

\section{The Petit--Quisquater--Tillich--Z\'{e}mor algorithm}\label{apx:PQTZ}

For the reader's convenience, we outline the generic algorithm of Petit, Quisquater, Tillich and Z\'{e}mor \cite{PQTZ09} for finding collisions for $q$ even. We describe their algorithm in a simplified language, generalize it to $p\ge 2$, and find optimal parameters:
collisions of length $\approx 12{\lg}^2 q$ in time $O(\sqrt{q}\lg q)$. This ignores the complexity of the second phase (discrete logarithms and LLL) of their attack, which we assume is smaller than $\sqrt{q}$. Setting their parameters so as to reduce
the running time below $O(\sqrt{q}\lg q)$ would render the length of the resulting collisions superpolynomial in $\lg q$.

\subsection{First phase: hashing into $\cT$}
The performance estimations of this phase can be proved, asymptotically, as in Section \ref{sec:rational}.
By Section \ref{subsec:T}, in time roughly $\sqrt{q}$ one can hash once into $\cT$ with a string of length
about $\lg q$.
Doing this $N:=\lg q$ times, we obtain (in time $\sqrt{q}\lg q$),
bitstrings $w_1,\dots,w_N$ each of length about $\lg q$ such that
$$h_A(w_1),\dots,h_A(w_N)\in\cT.$$

\subsection{Second phase: hashing into $\cK$}
Denote by $\lambda_1,\ldots ,\lambda_N$ the upper left entries of
$h_A(w_1),\dots,\allowbreak h_A(w_N)$, respectively.

Computing $N$ discrete logarithms in $\bbF_q$ and using the LLL algorithm,
find nonnegative integers $k_1,\dots,k_N$, with $\sqrt{k_1^2+\ldots+k_N^2}$ as small as possible,
such that
$$\lambda_1^{k_1}\ldots \lambda_N^{k_N}=1.$$
Taking all possibilities $k_i\in\{0,1\}$, $\lambda_1^{k_1}\ldots \lambda_N^{k_N}$ takes about
$$2^N\approx 2^{\lg q}=q$$
values. Thus, it is expected (although, thus far, unproved)
that the solution returned by the LLL algorithm satisfies
$$\sqrt{k_1^2+\ldots+k_N^2}\approx \sqrt{N}\approx \sqrt{\lg q}.$$
Let
$$v = w_1^{k_1}w_2^{k_2}\ldots w_N^{k_N},$$
where exponentiation denotes string concatenation.
By the Cauchy--Schwartz inequality,
$$|v| = k_1\cdot |w_1|+\ldots +k_N\cdot |w_N|  \le \sqrt{k_1^2+\ldots+k_N^2}\cdot \sqrt{{|w_1|}^2+\ldots+{|w_N|}^2},$$
with the right hand side being $\approx \sqrt{N}\cdot\sqrt{N{\lg}^2 q}=N\lg q= {\lg}^2q$.

Now,
$$h(v)\mx{1\\0} = h(w_1)^{k_1}\ldots h(w_n)^{k_n}\mx{1\\0} =
\lambda_1^{k_1}\ldots \lambda_n^{k_n}\mx{1\\0}=\mx{1\\0},$$
that is, for some $\beta\in\bbF_{q}$,
$$h(v)=\mx{1 & \beta\\0 &1}.$$
As $q=p^n$, we have that
$$h(v^p) = \mx{1 & p\beta\\0 &1} = \mx{1 & 0\\0 & 1},$$
that is, $v^p$ and the empty message hash to the same value.
We have that
$$|v^p| = p\cdot |v| \approx p{\lg}^2 q.$$
This completes our description of the Petit--Quisquater--Tillich--Z\'{e}mor algorithm.

Note that $p$ may be exponential in the security parameter.
To obtain shorter collisions, note that in the definition of $v$, if $u$ is obtained by any permutation of the order of the
$k_1+\ldots+k_N$ subwords $w_i$ in the word $v=w_1^{k_1}\ldots w_N^{k_N}$,
we still have by the same argument that
$$h(u)=\mx{1 & \gamma\\0 &1}$$
for some $\gamma\in\bbF_q$. Thus, \emph{$h(v)$ commutes with $h(u)$}, and we arrive at the collision
$$h(uv) = h(u)h(v) = h(v)h(u) = h(vu),$$
whose length is about $2\lg ^2 q$. Moreover, assuming for example that $k_1$ and $k_2$ are nonzero,
let
$$w=w_1^{k_1-1}w_2^{k_2-1}\ldots w_N^{k_N}$$
Taking $v = w_1w_2w$ and $u = w_2w_1w$, we know that
$$h(v w_2w_1)h(w)=h(vu)=h(uv)=h(u w_1w_2)h(w),$$
and therefore
$$h(v w_2w_1)=h(u w_1w_2),$$
a collision of length roughly
$$|v|+|w_1|+|w_2|\approx {\lg}^2 q+2\lg q \approx {\lg}^2 q.$$

The algorithms presented in Section \ref{sec:generic} are faster, and provide shorter collisions.
We stress that, unlike the Petit et al.\ algorithm, our generic algorithm
does not provide bitstrings hashing to the identity matrix (see, however, Remark \ref{rem:hashid}).

\section{The impossibility of palindromic collisions for $p>2$}\label{apx:nopali}

Let $q=2^n$ and let $\alpha$ be a primitive element of $\bbF_{2^n}$.
Let
\[
A_0=\mx{
        \alpha & -1 \\
        1 & 0 },\;\;
A_1=\mx{
        \alpha +1 & -1\\
        1 & 0}.
\]
Grassl, Ili\'{c}, Magliveras, and Steinwandt \cite{GIMS11} provide, in this case,
an efficient algorithm for finding palindromes $v\in \V$ of length $2n$ such that
the palindromes $0v0$ and $1v1$ hash to the same value under $h_A$.
This implies that the proposal in \cite{TZ2} is insecure.
Grassl et al.'s method does not generalize in any conceivable way to odd prime powers $q$.
In fact, we show here that for $q$ odd there are \emph{no} palindromes $v$ such that $0v0$ and $1v1$ form a collision.
Throughout, we write $h$ for $h_A$.

\bprp\label{noncol}
Let $v\in \V$ be a palindrome.
Then
\begin{enumerate}
\item $h(v)$ is of the form $\mx{ a & b \\ -b & d }$.

\item $h(0v0)-h(1v1)=\mx{ -2a\alpha -a - 2b & a \\ -a & 0 }$.

\item If $p>2$ then $h(0v0)\neq h(1v1)$.
\end{enumerate}
\eprp
\bpf
(1) We proceed by induction on the length of $v$.
The induction base consists of $|v|=1$ and $|v|=2$.
If $|v|=1$, i.e., $v=\beta\in\{0,1\}$, then
$$h(v) = \mx{
        \alpha+\beta & -1 \\
        1 & 0 }
$$
has the desired form.

Note that by direct calculation,

\begin{equation}
\label{theeq}
\mx{\alpha+\beta & -1 \\ 1 & 0 }\mx{ a & b \\ -b & d } \mx{ \alpha+\beta & -1 \\ 1 & 0 }
=  \mx{ a{(\alpha + \beta)}^2 + 2b(\alpha + \beta ) - d & -(a\alpha +a\beta +b) \\ a\alpha +a\beta +b & -a }.
\end{equation}

By Equation (\ref{theeq}), we have in particular (for $a=d=1$, $b=0$) that, for each $\beta\in\{0,1\}$,
$$h(\beta\beta)= A_\beta I A_\beta$$
has the desired form. This completes the verification of the induction base.

Induction step: assume that
$$h(v)=\mx{ a & b \\ -b & d }.$$
Then by Equation \eqref{theeq}
$h(\beta v\beta)=A_\beta h(v) A_\beta$ has the desired form for each
$\beta\in\{0,1\}$.

(2,3) Since $v$ is a palindrome, we have by the above calculation that
\[
h(0v0)-h(1v1)=\mx{ -2a\alpha -a - 2b & a \\ -a & 0 }.
\]
Hence, for $h(0v0) = h(1v1)$ to hold, $a$ must be $0$. This in turn implies that $2b=0$, which for $p>2$ implies
that $b=0$. Thus, $1=\det(h(v))=ad+b^2=0$, a contradiction.
\epf

Interestingly, it is pointed out in  \cite{GIMS11} that,
for $q$ a power of $2$ and a palindrome $v$, $h(0v1)=h(1v0)$ is equivalent to $h(0v0)=h(1v1)$.
For $q$ odd, we proved that $h(0v0)=h(1v1)$ is impossible (Proposition \ref{noncol}), but
that $h(0v1)=h(1v0)$ is provably possible (Theorem \ref{thm:lincol})!

\end{document}